# Investigation on the inclination angle of undrained shear slip surface in saturated soils based on mixture theory

**Hu Yayuan**[1,2] **Zhou Jian**[1,2]

(1. Research Center of Coastal and Urban Geotechnical Engineering, Zhejiang University, Hangzhou 310058, China; 2 Zhejiang Key Laboratory of the Development and Utilization of Underground Space, Zhejiang University, Hangzhou 310058, China)

**Abstract**: The inclination angle of the undrained shear slip surface in saturated soils is analyzed based on mixture theory. First, starting from the property that the bulk strain of soil skeleton is equal to the flow ratio of water discharged from soil skeleton, the energy conservation equation of saturated soil is obtained. According to state variables of energy equation and non-equilibrium thermodynamics, the mechanical mechanism underlying effective stress principle is revealed that Gibbs free energy of saturated soil is only expressed as a function of effective stress under isothermal process. Consequently, the deformation and strength of saturated soil are uniquely determined by the effective stress and not directly related to Newton's equilibrium equations, which governs the movements of solid-fluid two-phase components. The instability of soil skeleton is related to the applied forces and requires analysis based on the Newtonian equilibrium condition. The interaction between the solid-water components and the equilibrium equation of solid component are investigated under two working conditions: when permeability tensor equals zero and when it equals infinity. Combined with the Mohr-Coulomb strength theory, the inclination angle of the undrained shear slip surface is explored under Rankine's passive earth pressure in saturated soil. The results indicate that, when the permeability equals infinity, the uncoupled hydro-mechanical analysis is recommended, and the inclination angle of slip surface is $45° + \varphi'/2$; when the permeability equals zero, the fully coupled hydro-mechanical analysis is recommended, and the inclination angle of slip surface is $45°$. The permeability of actual saturated soil falls between the two extremes, the inclination angle of slip surface must be analyzed on a case-by-case basis.

## 0 Introduction

The analysis of the inclination angle of the undrained shear slip surface in saturated soils involves the mechanical mechanisms of the separate water and soil pressure calculation versus the combined water and soil pressure calculation, which is one of the challenging theoretical and engineering problems in soil mechanics. Li et al[1] and Li and Yu[2]. Davis et al[3], Zheng et al[4], Bolton et al[5], and Song & Lin[6] have conducted in-depth studies on this topic based on the saturated soil mechanics and the Mohr-Coulomb strength theory, which have provided great inspiration to the authors. However, due to the imperfection and lack of clarity in the mechanism foundations of saturated soil mechanics, currently, there are some divergences regarding the inclination angle of the undrained shear slip surface. Some scholars, starting from the principle of effective stress, believe the inclination angle is $45° + \varphi'/2$, while others, starting from the principle of total stress, believe the inclination angle is $45°$. Since the soil mechanics is very complex, a proper grasp of the fundamental concepts play a crucial role in theoretical analysis and engineering applications. Compared with other branches of applied mechanics, soil mechanics need be built upon clear, intuitive and immediately comprehensible concepts. To understand profoundly the basic concepts and mechanical mechanisms of soil mechanics from multiple angles, this paper attempts to investigate this difficult problem based on mixture theory.

## 1 Fundamentals of Mixture Theory

According to the exposition of mixture theory by the German rational mechanics scholar Professor Truesdell[7], saturated soil can be divided into two independent components: the solid phase and the fluid phase. Each component can be isolated from the mixture and treated as an independent continuous medium, provided that the influence of the other component is applied to the component in an appropriate action for studying. In soil mechanics, the concerns in engineering field focus on the deformation and stability of solid phase. Therefore, the solid phase is usually isolated from the mixture and treated as an independent continuous medium in mixture analysis, while the influence of the fluid phase on the solid phase is accounted for by applying the fluid-solid interaction force to the solid phase. In practice, the solid and fluid phases in saturated soils exhibit two distinct configurations at different scales[7]. One is the component physically real microscale configuration, such as solid particles and porous fluid in saturated soil. In this configuration, the space occupied by the solid phase cannot be occupied by the fluid phase, and vice versa. In the mixture theory, the strains they produce are called component matrix strains. In the mixture theory, the matrix of solid component and fluid component are usually assumed to be zero, that means the component matrix deformation of the fluid and solid phases is not considered. The other is the macroscopic configuration where component is averaged by volume fraction and vary continuously. The strain produced in this configuration is called component strain in the terminology of classical mixture theory. The solid phase strain is denoted by $\varepsilon_S$, and the fluid phase strain by $\varepsilon_F$. In engineering mixture theory, the solid phase strain caused by the variation in the void ratio of porous media is called soil skeleton strain. In saturated soil, since the solid component matrix strain caused by soil particles is not considered, the solid strain is equal to the soil skeleton strain. Thus, in saturated soil mechanics, solid deformation is often referred to as soil skeleton deformation.

For brevity, the following notation is adopted: the subscript $S$ denotes the solid phase and the subscript $F$ denotes the fluid phase, and $\alpha = \{S, F\}$ is used as the constituent index. Let $\rho_{R\alpha}$ be the material density (also called real density in saturated soil mechanics) of component $\alpha$, $\rho_\alpha = n_\alpha \rho_{R\alpha}$ be the average density of component $\alpha$, and $n_\alpha$ be the volume fraction. For saturated soil, the volume fractions $n_\alpha$ satisfy:

$$n_S + n_F = 1 \tag{1}$$

Let $\boldsymbol{\sigma}_S$ be the stress tensor borne by the solid component in saturated soil, and $\boldsymbol{\sigma}_{Fm}$ be the spherical stress borne by the fluid component; let $P_S = (\boldsymbol{\sigma}_S : \mathbf{I})/(3n_S)$ be the solid matrix pressure; let $P_F = \sigma_{Fm}/n_F$ be the fluid matrix pressure or pore pressure; and let $\boldsymbol{\sigma}$ be the total stress tensor of the saturated porous medium. All are positive in compression. According to engineering mixture theory, there are[8]:

$$\boldsymbol{\sigma} = \boldsymbol{\sigma}_S + \sigma_{Fm}\mathbf{I} = \boldsymbol{\sigma}_S + (n_F P_F)\mathbf{I} \tag{2}$$

It is worth noting that Equation (2) is not the effective stress formula defined by Terzaghi.

Let $x_S$ and $x_F$ be the current configuration coordinates of the fluid and solid phases respectively. According to mixture theory, the Newton's second laws (That is, force equilibrium conditions) satisfying the solid and fluid components are[8]:

$$\nabla \cdot (-\boldsymbol{\sigma}_S) + n_S \rho_{SR} g - \hat{\boldsymbol{p}} = n_S \rho_{SR} \frac{d_S^2 x_S}{dt^2}, \tag{3}$$

$$\nabla(-n_F P_F) + n_F \rho_{RF} g + \hat{\boldsymbol{p}} = n_F \rho_{RF} \frac{d_F^2 x_F}{dt^2}. \tag{4}$$

In the equations, $\hat{\boldsymbol{p}}$ is the interaction term between the fluid-solid phases. Let $x$ be the current configuration coordinate of the mixture, and $\rho_{sat} = n_S \rho_{SR} + n_F \rho_{RF}$ be the saturated density of the saturated soil. We have:

$$\frac{dx}{dt} = \frac{n_S \rho_{SR}}{\rho_{sat}} \frac{d_S x}{dt} + \frac{n_F \rho_{RF}}{\rho_{sat}} \frac{d_F x}{dt} \tag{5}$$

Let $u_S = (d_S x_S/dt) - (dx/dt)$ and $u_F = (d_F x_F/dt) - (dx/dt)$, then Newton's second law for the entire mixture is[9]:

$$\nabla \cdot (-\boldsymbol{\sigma}_S - n_F P_F \mathbf{I} + n_S \rho_{RS} u_S \times u_S + n_F \rho_{RF} u_F \times u_F) + \rho_{sat} g = \rho_{sat} \frac{d^2 x}{dt^2} \tag{6}$$

Using Equation (2), Equation (6) can be transformed into:

$$\nabla \cdot (\boldsymbol{\sigma} + n_S \rho_{RS} u_S \times u_S + n_F \rho_{RF} u_F \times u_F) + \rho_{sat} g = \rho_{sat} \frac{d^2 x}{dt^2} \tag{7}$$

The reason why Equation (7) differs from the commonly used Newton's second law for overall saturated soil mixtures in soil mechanics is that the solid and fluid phases are considered as two mutually independent continuous media in the mixture theory. Different component stresse act on different continuous media; therefore, the stresses acting on these two different continuous media cannot be simply added, instead, additional stress terms associated with the relative velocity between the two independent continuous media must be included.

In soil mechanics, the primary focus lies in the deformation and stability of solid phase, and thus warrants in-depth investigation. The following analysis is based on the solid configuration unless otherwise specified. For convenience in writing, let $\dot{A} = d_S A/dt$. According to engineering mixture theory, $\dot{\varepsilon}_S = \nabla \cdot w$ is valid when the deformation of the fluid and solid matrices is not considered. Consequently, the law of conservation of energy for saturated soil is[8]:

$$\dot{U} = \boldsymbol{\sigma}': \dot{\boldsymbol{\varepsilon}}_S + \boldsymbol{W}_F \cdot (P_F \nabla n_F - \hat{\boldsymbol{p}}) \tag{8}$$

In the equation, $\boldsymbol{W}_F = (d_F \boldsymbol{x}_F / dt) - (d_S \boldsymbol{x}_S / dt)$. $U$ is the internal energy, which is a function of solid strain $\boldsymbol{\varepsilon}_S$ and entropy $\eta$, i.e., $U(\boldsymbol{\varepsilon}_S, \eta)$. Taking the total differential of $U(\boldsymbol{\varepsilon}_S, \eta)$ and substituting it into Equation (8) gives:

$$(\boldsymbol{\sigma}' - \frac{\partial U}{\partial \boldsymbol{\varepsilon}_S}) : \dot{\boldsymbol{\varepsilon}}_S + \frac{\partial U}{\partial \eta} \dot{\eta} + \boldsymbol{W}_F \cdot (P_F \nabla n_F - \hat{\boldsymbol{p}}) = 0 \tag{9}$$

Set $\theta$ as the temperature. From thermodynamics, it follows that:

$$\theta = \frac{\partial U}{\partial \eta} \tag{10}$$

Note that the state variable $\boldsymbol{\varepsilon}_S$ can vary free while the other variables remain unchanged, hence, from Equation (9) we obtain[9]:

$$\boldsymbol{\sigma}' = \frac{\partial U}{\partial \boldsymbol{\varepsilon}_S} \tag{11}$$

Substituting Equation (10)-(11) into Equation (9) and applying the second law of thermodynamics yields:

$$\theta \dot{\eta} = \boldsymbol{W}_F \cdot (\hat{\boldsymbol{p}} - P_F \nabla n_F) \geq 0 \tag{12}$$

Let the Gibbs free energy $G(\boldsymbol{\varepsilon}_S, \theta) = \boldsymbol{\sigma}' \boldsymbol{\varepsilon}_S + \theta \eta - U_S(\boldsymbol{\varepsilon}_S, \eta)$, i.e, two Legendre transformation is performed from internal energy. After a derivation similar to that in Reference [8], the following equations are obtains as following under isothermal conditions (where $\theta$ is constant and can be omitted)[8]:

$$\boldsymbol{\varepsilon}_S = \frac{\partial G(\boldsymbol{\sigma}')}{\partial \boldsymbol{\sigma}'} \tag{13}$$

$$\hat{\boldsymbol{p}}_F - P_F \nabla n_F = \frac{\partial \phi(\boldsymbol{W}_F)}{\partial \boldsymbol{W}_F} \tag{14}$$

Equation (13) is the principle of effective stress for saturated soil. This equation indicates that the mechanical mechanism of the effective stress principle is that the deformation energy of the soil skeleton in saturated soil can only be expressed as a function of effective stress. That is, the effective stress is used to establish the stress-strain relationship based on strain deformation energy. Therefore, effective stress belongs to the mechanical category of deformation mechanics established by Hooke and is only defined through strain energy. It has indeed no direct relationship with the component force equilibrium equations (3)~(4) established by Newton's second law[10-11]. If the effective stress principle is regarded as a consequence of Newton's force equilibrium condition, it may readily lead to a fundamental conceptual conflict between deformation mechanics and Newton's mechanics in soil mechanics. According to the generalized Hooke's law, the elastic constitutive equation can be set as $G(\boldsymbol{\sigma}') = \boldsymbol{\sigma}' : \boldsymbol{\Gamma} : \boldsymbol{\sigma}'/2$, where $\boldsymbol{\Gamma}$ is the fourth-order compliance tensor[8]. And the seepage equation can be set as $\phi(\boldsymbol{W}_F) = \boldsymbol{W}_F \cdot \boldsymbol{K}^{-1} \cdot \boldsymbol{W}_F /2$, where $\boldsymbol{K}$ is the second-order permeability tensor[8]. Then we have:

$$\boldsymbol{\varepsilon}_S = \boldsymbol{\Gamma} : \boldsymbol{\sigma}' \qquad (15)$$

$$\hat{\boldsymbol{p}} - P_F \nabla n_F = \boldsymbol{K}^{-1} \cdot \boldsymbol{W} \qquad (16)$$

The instability of saturated soil is related to force equilibrium and determined by the force equilibrium equation (3) and (4) and (6), in which the force equilibrium and instability of soil skeleton is determined by equation (3). When forces and strength on a certain plane of soil skeleton in saturated soil reaches Newton's equilibrium condition i.e. the limit equilibrium state, the soil skeleton will slide along the failure surface. From the perspective of mixture theory, the soil skeleton and the pore fluid are treated as two independent continuous media. Each force equilibrium is influenced not only by their respective component stresses but also by the soil-fluid interaction $\hat{\boldsymbol{p}}$. When the soil skeleton slides along the failure surface, the resulting motions of the fluid and the soil skeleton do not depends solely on the respective stress of their component but also on the magnitude and nature of $\hat{\boldsymbol{p}}$. Two limiting cases are discussed in detail below.

## 2 Solution for $\boldsymbol{K} = \infty$

Substituting $\boldsymbol{K} = \infty$ into Equation (16), we have:

$$\hat{\boldsymbol{p}} = P_F \nabla n_F \qquad (17)$$

At this time, the drag force of the soil skeleton on pore water is zero, and the coupling effect between the fluid and solid component is only made of this term. Considering that elastoplastic failure analysis is a static analysis, acceleration is not considered. Taking acceleration as zero, Equations (3)~(4) become:

$$\nabla \cdot (-\boldsymbol{\sigma}_S) + n_S \rho_{SR} g - \hat{\boldsymbol{p}} = 0 ; \qquad (18)$$

$$\nabla(-n_F P_F) + n_F \rho_{RF} g + \hat{\boldsymbol{p}} = 0 \cdot \qquad (19)$$

Substituting Equation (17) into Equation (19) yields:

$$\nabla P_F - \rho_{RF} g = 0 \qquad (20)$$

Substituting Equations (2) and (17) and into (19) Equation (18) yields:

$$\nabla \cdot (\boldsymbol{\sigma} - P_F \boldsymbol{I}) - n_S \rho_{SR} g + n_S \nabla P_F = 0 \qquad (21)$$

Using the effective stress formula $\boldsymbol{\sigma}' = \boldsymbol{\sigma} - P_F \boldsymbol{I}$ and substituting Equation (20) into Equation

(21), we get:

$$\nabla \cdot \boldsymbol{\sigma}' - (n_S \rho_{SR} + n_F \rho_{FR} - \rho_{FR})g = \nabla \cdot \boldsymbol{\sigma}' - (\rho_{sat} - \rho_{FR})g = \nabla \cdot \boldsymbol{\sigma}' - \rho'g = 0 \tag{22}$$

The derivation of Equation (22) makes use of the equation that the buoyant unit weight of saturated soil is equal to $\rho'g = (\rho_{sat} - \rho_{FR})g$. Equation (22) reflects the skeleton force equilibrium equation expressed in terms of effective stress. Equation (22) indicates that, in this case, the force equilibrium analysis of the soil skeleton can be equivalently treated as a single-phase analysis in which the effective stress $\boldsymbol{\sigma}'$ acts as the internal force and the buoyant unit weight $\rho'g$ acts as the external body force.

Now consider the Rankine passive earth pressure problem in this condition under undrained conditions. From Equation (22), the vertical stress at depth z is $\sigma'_v = \rho'gh$, and the horizontal stress is $\sigma'_h = K_0 \rho'gh$, where h is depth and $K_0$ is horizontal earth pressure coefficients. Assuming the effective cohesion of Mohr–Coulomb strength for saturated soil be $c'$ and the effective frictional angle $\varphi'$. The Rankine passive earth pressure at depth $h$ is $\sigma'_p$. Under undrained conditions, the excess pore pressure can be obtained as $P_F = (\sigma_p - K_0 \rho'gh)/2$. Hence, the horizontal effective stress is $\sigma'_1 = \sigma_p/2 + K_0 \rho'gh/2$, and the vertical effective stress is $\sigma'_3 = (1 + K_0/2)\rho'gh - \sigma_p/2$. When the horizontal stress reaches the Rankine passive earth pressure, the soil element is at the limit-equilibrium state. In the effective-stress Mohr circle, the shear stress equals the shear strength on the plane at an angle of $\alpha = 45° + \varphi'/2$ to the first principal stress, and this means the soil element is in the force limit-equilibrium state on this plane. Connecting the slip planes of all soil elements forms the Rankine passive earth-pressure slip surface. On this surface, the strength and forces reach Newton force equilibrium condition, that is, force limit equilibrium state. Consequently, the inclination of Rankine passive earth-pressure surface is equal to the force limit equilibrium plane of soil element at $\alpha = 45° + \varphi'/2$. According to the trigonometric relationship among the Mohr-circle radius, the horizontal coordinate of the circle center and the cohesion, we obtains $\sigma'_p = \rho'gh + 2c'\cos\varphi' + (1+K_0)\rho'gh\sin\varphi'$, and hence the Rankine passive earth pressure is given by:

$$E' = \frac{1}{2}\rho'gh^2 + \frac{1+K_0}{2}\rho'gh^2 \sin\varphi' + 2c'h\cos\varphi' \tag{23}$$

The passive pressure generated by the fluid is:

$$E_F = \frac{1}{2}\rho_{FR}gh^2 \tag{24}$$

The total passive earth pressure resulting from the sum of both is:

$$E = \frac{1}{2}\rho gh^2 + \frac{1+K_0}{2}\rho' gh^2 \sin\varphi' + 2c'h\cos\varphi' \qquad (25)$$

Equations (23)-(25) indicate that, in this case, the earth-pressure analysis is appropriately treated using the uncoupled hydro-mechanical analysis.

## 3 Solution for $K = 0$

Substituting $K = 0$ into Equation (16), and using the definition of $W_F$, we have:

$$(d_F x_F / dt) - (d_S x_S / dt) = W_F = 0 \qquad (26)$$

Equation (26) indicates that, at this moment, in addition to the solid-fluid interaction reflected by effective stress, there also exists a drag force from the soil skeleton on the pore water. Moreover, the drag force between the fluid and solid phases is so large that no velocity difference occurs between the solid and fluid phases. According to Equation (5) and Equation (26), we get:

$$\frac{d_F x_F}{dt} = \frac{d_S x_S}{dt} = \frac{dx}{dt}, \quad \frac{d^2 x}{dt^2} = \frac{d_F^2 x_F}{dt^2} = \frac{d_S^2 x_S}{dt^2} \qquad (27)$$

Using Equation (3), Equation (5) and Equation (28), and noting that $u_S = u_F = 0$ according to Equation (27), we can obtain:

$$\nabla \cdot (-\boldsymbol{\sigma}_S) + n_S \rho_{SR} g - \hat{\boldsymbol{p}} = n_S \rho_{SR} \frac{d_S^2 x_S}{dt^2} = n_S \rho_{SR} \frac{d^2 x}{dt^2} = \frac{n_S \rho_{SR}}{\rho_{sat}} (\nabla \cdot \boldsymbol{\sigma} + \rho g) \qquad (28)$$

From Equation (28), we get:

$$\hat{\boldsymbol{p}} = -\nabla \cdot (\boldsymbol{\sigma}_S) - \frac{n_S \rho_{SR}}{\rho_{sat}} \nabla \cdot \boldsymbol{\sigma} \qquad (29)$$

Still taking the soil skeleton as the research object, noting that elastoplastic failure analysis is a static problem so that acceleration is not considered. Taking acceleration as zero, Equation (3) becomes:

$$\nabla \cdot (-\boldsymbol{\sigma}_S) + n_S \rho_{SR} g - \hat{\boldsymbol{p}} = 0 ; \qquad (30)$$

Substituting Equation (29) into Equation (30), we have:

$$\nabla \cdot \boldsymbol{\sigma} - \rho_{sat} g = 0 \qquad (31)$$

Equation (31) indicates that when $K = 0$, the force equilibrium analysis of the soil skeleton can be equivalently treated as a single-phase analysis in which the total stress $\boldsymbol{\sigma}$ acts as the internal force and the saturated unit weight $\rho_{sat} g$ acts as the external body force.

Now consider the Rankine passive earth-pressure problem under this condition. From Eq. (31), the vertical total stress at depth $h$ is $\sigma_v = \rho gh$ and the horizontal total stress is $\sigma_h = K_0 \rho gh$. According to the derivation by Song and Lin[6], the undrained strength of saturated soil can be expressed as:

$$c_u = (\sigma_1 - \sigma_3)/2 = c'\cos\varphi' + 0.5(1+K_0)\sin\varphi'\rho' gh \qquad (32)$$

Let the horizontal Rankine passive earth pressure at depth $h$ be $\sigma_p$. Since the undrained shear strength of saturated soil is independent of the total stress, the Mohr–Coulomb failure envelope is a straight line parallel to the horizontal axis. When the soil element reaches the limit-equilibrium state under the action of the horizontal Rankine passive earth pressure, the total-stress Mohr circle intersects the Mohr–Coulomb failure envelope at the point of maximum shear stress. In this case, the slip plane is oriented at an angle $\alpha = 45°$ to the first principal stress, where the shear stress equals the shear strength and the soil element is at force limit equilibrium state. Connecting the slip planes of all elements forms the Rankine passive earth-pressure slip surface. On this surface, the strength and forces reach Newton force equilibrium condition, that is, the force limit equilibrium state. Consequently, the inclination of the slip surface is equal to $\alpha = 45°$. Since the radii of the undrained Mohr circles are the same for saturated soil, it follows that $\sigma_p = \rho g h + 2c_u$. Substituting Equation (32) into $\sigma_p = \rho g h + 2c_u$ and integrating with respect to depth yields the Rankine passive earth pressure:

$$E = \frac{1}{2}\rho g h^2 + \frac{1+K_0}{2}\rho' g h^2 \sin\varphi' + 2c'h\cos\varphi' \qquad (33)$$

Equations (31)-(33) indicate that, in this case, the fully coupled hydro-mechanical method (the total stress method) is appropriate for the earth-pressure analysis.

## 4 Solution for A General $K$

In practice, the permeability tensor $K$ of soils falls between zero and infinity. Under this case, when $W = 0$, the interactive force $\hat{p}$ between the solid and fluid phase is the same as that obtained in the "Solution for $K = \infty$" section. Consequently, the stress analysis of the solid soil skeleton under this condition is identified to the analysis presented in" "Solution for $K = \infty$" section. However, it is worth noting that undrained condition for saturated soil is typically only reflected in a boundary condition. The internal behavior of soil can only satisfy $\nabla \cdot W_F = 0$ under homogeneous circumstance, and even then, achieving $W_F = 0$ is not guaranteed. Therefore, when the soil skeleton reaches the limit equilibrium state and become unstable, forming a velocity discontinuous surface, the internal homogeneity of saturated soil is lost. Under this condition, the internal uniformity no longer holds. Consequently, in general, the undrained boundary condition cannot guarantee that the internal soil satisfy $\nabla \cdot W_F = 0$. Under this case, on the one hand, the mutually independent solid and fluid phases inevitably tend to separate in velocity, generating seepage drag force between the solid and fluid phases and driving their motion together. As a result, the coupled hydro-mechanical response between the solid and fluid phases tends to approach the fully coupled hydro-mechanical regime. On the other hand, Instability in the soil skeleton can induce misalignment at micro- and meso-structure level. Consequently, Change within the soil mass, such as the deformation of void, tensile cracking, dilation and bulging facilitate separation and drainage of fluid from the solid skeleton. This results in localized consolidation, driving the coupled hydro-mechanical response between the solid and fluid phases toward to approach the uncoupled hydro-mechanical regime. Due to the coupled hydro-mechanical responses of the solid and fluid phases are highly complex, it is difficult to quantitatively determine the inclination angle of the undrained shear slip surface in saturated soil. It is only possible to qualitatively conclude that the inclination of the undrained shear slip surface falls between $\alpha = 45°$ and $\alpha = 45° + \varphi'/2$. Generally speaking, for saturated sandy soils with relatively high permeability, the tests conducted by Li and co-workers[2] indicate that the slip-surface inclination under undrained shearing approaches $\alpha = 45° + \varphi'/2$. In this case, the earth pressure and slope stability can be approximated analyzed using the uncoupled hydro-mechanical method[2]. for saturated clayey soils with low permeability. The conventional triaxial CU strength parameters, or using the undrained total shear strength $S_u$ in $\varphi = 0$ method are often used as fully coupled hydro-mechanical methods to analyze the earth pressure and slope stability in geotechnical engineering[2].

## 5 Conclusion

(1) The mechanical mechanism of effective stress principle for saturated soils is that the Gibbs free energy must be expressed in terms of the effective stress, so the effective stress concept belong to the category of deformation mechanics. The deformation and strength of soil skeleton in saturated soil is determined by effective stress. The force equilibrium and stability of skeleton is determined by Newton equilibrium condition. The strength of skeleton as a force influence the Newton equilibrium condition.

(2) When the permeability tensor of saturated soil tends to infinity, the uncoupled hydro-mechanical analysis is appropriate, and the inclination of the undrained shear slip surface is

$\alpha = 45° + \varphi'/2$

(3) When the permeability tensor of saturated soil tends to zero, the fully coupled hydro-mechanical analysis is required, and the inclination of the undrained shear slip surface is $\alpha = 45°$

(4) The practical permeability tensor of saturated soil falls between zero and infinity. Due to the coupled hydro-mechanical responses of the solid and fluid phases are highly complex, it is difficult to quantitatively determine the inclination angle of the undrained shear slip surface in saturated soil. It is only possible to qualitatively conclude that the inclination of the undrained shear slip surface falls between $\alpha = 45°$ and $\alpha = 45° + \varphi'/2$.